\documentclass{jps-cp}
\usepackage{txfonts} %Please comment out this line unless the txfonts package is available in your LaTeX system.
\usepackage{xcolor}
\usepackage{booktabs}
\usepackage{multirow}
\usepackage{graphicx}

\title{Effect of Correlation on the Pressure-Induced Insulator to Metal Transition in Complex Spin-Orbit Coupled Double Perovskite La$_2$CoIr$_{1-x}$Ti$_x$O$_6$}

\author{Vineeta \textsc{Yadav}, Sromona \textsc{Nandi} and Rudra Sekhar \textsc{Manna}$^{*}$}

\inst{Department of Physics, Indian Institute of Technology Tirupati, Andhra Pradesh 517619, India\\
}

\email{$^*$rudra.manna@iittp.ac.in}

\recdate{September 22, 2026}

\abst{We study the effect of electron correlation ($U$) and hydrostatic pressure on the electronic properties of double perovskites La$_2$CoIr$_{1-x}$Ti$_x$O$_6$ for $x$ = 0, 0.5, and 1, using first-principles electronic structure calculations within density functional theory (DFT). At ambient pressure, we systematically vary the electron correlation, i.e., on-site Coulomb interaction to study its effect on band gap and structural parameters. Increasing the value of $U$ enhances the band gap of all three compositions while producing only a slight variation in structural parameters. The effect of hydrostatic pressure is also studied for the same compositions. The estimated band gap values at ambient pressure closes at approximately 39, 42, and 53 GPa for $x$ = 0 (for $U_\mathrm{Co}$ = 5 eV and $U_\mathrm{Ir}$ = 2 eV), 0.5 (for $U_\mathrm{Co}$ = 5 eV and $U_\mathrm{Ir}$ = 2 eV), and 1 (for $U_\mathrm{Co}$ = 4 eV), respectively, indicating pressure-induced insulator-to-metal transitions. %Further, for $x$ = 1, we studied the insulator-to-metal transition for $U_\mathrm{Co}$ = 2, 3, and 4 eV, where the critical pressure required to get the metallic phases increases from 42 to 47 and 53 GPa, respectively. 
These results show that electron correlation stabilises the insulating state, whereas hydrostatic pressure favours the metallic state by closing the band gap. These findings highlight the tunability of correlated oxides through electron correlations and pressure, offering insights for materials design.}

\kword{Electron correlation, hydrostatic pressure, double perovskites}

\begin{document}
\maketitle

\section{Introduction}

Double perovskites (DPs), with the general formula A$_2$BB$^\prime$O$_6$, have garnered considerable attention over the past few decades due to their distinctive electronic and magnetic properties, which arise from the intricate interplay among charge, spin, orbital, and lattice degrees of freedom~\cite{Vasala2015}. These systems are also known to exhibit Mott insulating behaviour due to the presence of finite on-site Coulomb repulsion ($U$). In addition, there is a delicate balance between on-site Coulomb repulsion ($U$), crystal electric field (CEF) effects, and spin-orbit coupling (SOC) originating from transition metal ions~\cite{Cavichini2018}. In this context, examining the $U$ dependency provides a direct way to understand how the strength of electron correlation affects the ground state properties of the oxides. A systematic variation of $U$ allows us to understand the evolution of electron localisation, band gap, local magnetic moment, and structural parameters. Such analysis is especially relevant when there is more than one magnetic cation present in the system, where the electronic properties may arise from the coupled response of both magnetic cations~\cite{Yasir2025}. While the variation of $U$ enhances electron localisation and stabilises the insulating nature of the system, hydrostatic pressure provides a complementary route to tune the electronic state by modifying the orbital overlap, bandwidth, and magnetic exchange interactions. Despite the rich physics exhibited by these double perovskite materials, the influence of hydrostatic pressure on the insulator-metal transition in DPs remains relatively less explored, especially for the antiferromagnetic systems. A recent report on La$_2$CoTiO$_6$ ($x$ = 1 hereafter), which showed an antiferromagnetic transition at 15 K, demonstrated that under hydrostatic pressure, the system transforms from an antiferromagnetic insulating (AFM-I) phase to an antiferromagnetic metallic (AFM-M) phase around 42 GPa and eventually reaches a nonmagnetic phase near 130 GPa~\cite{Nandi2024}. In this work, we examine how varying the correlation strength ($U$) affects the electronic properties of La$_2$CoIr$_{1-x}$Ti$_x$O$_6$ for $x$ = 0, 0.5, and 1 at ambient pressure. We further study the pressure dependence of the band gap for $x$ = 0, 0.5, and 1. In addition, we also study how increasing electron correlation ($U$) affects the pressure evolution of the band gap for $x$ = 1 and the requirement of critical pressure for the AFM-I to AFM-M transition.

\section{Computational Details}

First-principles calculations based on density functional theory (DFT) were carried out using the Vienna $ab$ $initio$ Simulation Package (VASP)~\cite{Kresse1996} employing spin-polarised methods within the projector augmented-wave (PAW) framework~\cite{Blochl1994,Kresse1999} and a plane-wave basis set. The Perdew-Burke-Ernzerhof (PBE)~\cite{PBE1996} functional within the generalized gradient approximation (GGA) was used for the exchange-correlation potential and a plane-wave cutoff energy of 500 eV. For $x$ = 0 and 0.5, a $\Gamma$-centred $6\times6\times4$ $k$-point mesh was employed, and the effects of both on-site Coulomb interaction and spin--orbit coupling were included through the GGA + $U$ + SOC approach, whereas for $x$ = 1,  the calculations were performed using GGA + $U$ without SOC, more details are available in~\cite{Nandi2024,Nandi2025}. All spin-polarised calculations were performed considering the collinear antiferromagnetic (AFM) configuration, based on the magnetic ground-state investigations reported in our previous studies~\cite{Nandi2024,Nandi2025}. The value of $U_{\mathrm{Co}}$ was systematically varied from 1--5 eV with a step size of 1 eV, while $U_{\mathrm{Ir}}$ was fixed at 2 eV and $U_{\mathrm{Ti}}$ was taken as zero owing to the nonmagnetic Ti$^{4+}$ electronic configuration. However, no noticeable change in the electronic properties was observed upon varying $U_{\mathrm{Ti}}$~\cite{Nandi2025}. For $x$ = 0 and 0.5, the Hund's exchange parameter ($J$) was kept fixed at $J_\textrm{Co}$ = 1.03 eV and $J_\textrm{Ir}$ = 0.94 eV~\cite{Nandi2025}. The internal positions of all atoms were thoroughly relaxed until the forces on each atom were reduced below 10$^{-3}$ eV/\AA, with the total energy convergence cutoff to 10$^{-7}$ eV. We use the experimental structure for $x$ = 0 and 1 in our theoretical calculations, whereas for $x$ = 0.5, we use the $x$ = 1 structure and replace one Ti atom with Ir. All three structures were subsequently relaxed at zero pressure before further calculations. After relaxation, static calculations were performed using the tetrahedron method with Bl\"{o}chl corrections to obtain the density of states (DOS). The hydrostatic pressure was applied uniformly by adjusting the scaling factor of the lattice vectors as mentioned in references~\cite{Nandi2024,Mei2009}, resulting in a reduction of the unit cell volume.

\section{Results and Discussions}

\begin{figure}[tbh]
\centering
\includegraphics[width=1.0\linewidth]{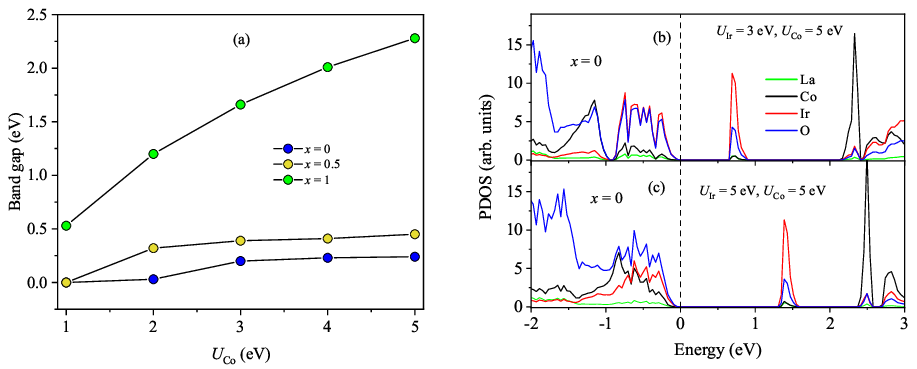}
\caption{(a) Variation of the band gap with the on-site Coulomb interaction ($U_\mathrm{Co}$) for $x$ = 0, 0.5 and 1, while $U_\mathrm{Ir}$ = 2 eV is kept fixed. For $x$ = 0, atom-projected partial density of states (PDOS) at fixed $U_{\mathrm{Co}}$ = 5 eV for (b) $U_{\mathrm{Ir}}$ = 3 eV and (c) $U_{\mathrm{Ir}}$ = 5 eV. The corresponding band gap values are 0.62 and 1.32 eV, respectively. With increasing $U_{\mathrm{Ir}}$, the Ir states show a pronounced shift away from the Fermi level, whereas the Co states remain comparatively less affected. This indicates that the electronic structure near the Fermi level is more sensitive to the correlation strength at the Ir site. The dashed vertical line marks the Fermi level ($E_{\mathrm{F}}$ = 0 eV).} \label{LCITO}
\end{figure} 

\subsubsection{La$_2$CoIrO$_6$ ($x$ = 0)}

The effect of the on-site Coulomb interaction, $U$, on the structural and electronic properties of $x$ = 0 was examined by varying $U_{\mathrm{Co}}$ from 1 to 5 eV in steps of 1 eV. When $U_{\mathrm{Ir}}$ was fixed at 1 eV, the system remained metallic throughout the investigated $U_{\mathrm{Co}}$ range (not shown). This indicates that increasing the correlation strength on Co alone is insufficient to open a gap, since the Ir-$5d$ states continue to contribute near the Fermi level. In contrast, for $U_{\mathrm{Ir}}=2$ eV, the system remains metallic at $U_{\mathrm{Co}}$ = 1 eV and develops only a very small gap at $U_{\mathrm{Co}}$ = 2 eV. A clear insulating state appears at $U_{\mathrm{Co}}$ = 3 eV, where the band gap increases almost linearly to 0.20 eV, followed by only a small further increase for $U_{\mathrm{Co}}$ = 4 and 5 eV, as shown in Fig.~\ref{LCITO}(a). As an additional test, $U_\mathrm{Co}$ was increased to 10 eV (not shown) while keeping $U_\mathrm{Ir}$ = 2 eV, for which the band gap increases only to about 0.34 eV. This reveals that the band gap is weakly affected by higher $U_\mathrm{Co}$ value. These results suggest that the gap formation arises from the combined effect of correlations on both the Co-3$d$ and Ir-5$d$ states. The lattice parameters change only slightly with increasing $U_{\mathrm{Co}}$, as shown in table~\ref{tab:U_paramters}. We have also examined the role of Ir in the electronic structure by fixing the $U_{\mathrm{Co}}$ = 5 eV and varying the $U_{\mathrm{Ir}}$ from 3 to 5 eV. The band gap increases from about 0.24 eV at $U_{\mathrm{Ir}}$ = 2 eV to 0.62 eV at 3 eV and 1.32 eV at 5 eV, showing a much stronger sensitivity to $U_{\mathrm{Ir}}$ than to $U_{\mathrm{Co}}$, shown in Fig.~\ref{LCITO}(b) and (c). The atom-projected partial density of states (PDOS) shows that increasing $U_{\mathrm{Ir}}$ shifts the unoccupied Ir-$d$ states away from the Fermi level, while the main Co-$d$ features remain comparatively unchanged (see Figs.~\ref{LCITO}(b) and (c)). This indicates that the Ir site plays a major role in controlling the band gap of $x$ = 0. The higher $U_\textrm{Ir}$ values were considered only to examine the sensitivity of the electronic structure, as excessively large $U$ values may overestimate the band gap and may not represent realistic interaction strengths. Overall, the results show that the transition from a metallic to a narrow-gap insulating state in $x$ = 0 is mainly due to changes in its electronic states, with both Co and Ir contributing to the gap formation.

\begin{table*}[ht]
\centering
\scriptsize
\caption{Calculated lattice parameters and band gap for $x$ = 0, 0.5, and 1. The calculations on the $x$ = 0 and 0.5 systems were performed using GGA + $U$ + SOC, whereas GGA + $U$ was used for $x$ = 1. For $x$ = 0 and 0.5, $U_{\mathrm{Ir}}$ was fixed at 2 eV, while $U_{\mathrm{Co}}$ was varied from 1 to 5 eV in steps of 1 eV.}
\label{tab:U_paramters}
\begin{tabular}{c c c c c c}
\hline
System & $U_\mathrm{Co}$~(eV) & $a$~(\AA) & $b$~(\AA) & $c$~(\AA) & $E_\mathrm{g}$~(eV) \\
% & (eV) & (\AA) & (\AA) & (\AA) & (eV) & ($\mu_{\mathrm{B}}$) & ($\mu_{\mathrm{B}}$) \\
\hline
         & 1 & 5.554 & 5.704 & 7.906 & 0.00  \\
         & 2 & 5.555 & 5.706 & 7.908 & 0.03  \\
$x$ = 0  & 3 & 5.559 & 5.710 & 7.913 & 0.20  \\
         & 4 & 5.559 & 5.710 & 7.914 & 0.23  \\
         & 5 & 5.563 & 5.714 & 7.918 & 0.24  \\
\hline
          & 1 & 5.562 & 5.672 & 7.851 & 0.00  \\
          & 2 & 5.635 & 5.639 & 7.953 & 0.32  \\
$x$ = 0.5 & 3 & 5.639 & 5.643 & 7.958 & 0.39  \\
          & 4 & 5.639 & 5.644 & 7.960 & 0.41  \\
          & 5 & 5.643 & 5.647 & 7.964 & 0.45  \\
\hline
         & 1 & 5.514 & 5.588 & 7.805 & 0.53   \\
         & 2 & 5.514 & 5.593 & 7.812 & 1.20  \\
$x$ = 1  & 3 & 5.514 & 5.596 & 7.818 & 1.66  \\
         & 4 & 5.515 & 5.598 & 7.822 & 2.01  \\
         & 5 & 5.515 & 5.599 & 7.825 & 2.28  \\
\hline
\end{tabular}
\end{table*}

\begin{figure}[tbh]
\centering
\includegraphics[width=1.0\linewidth]{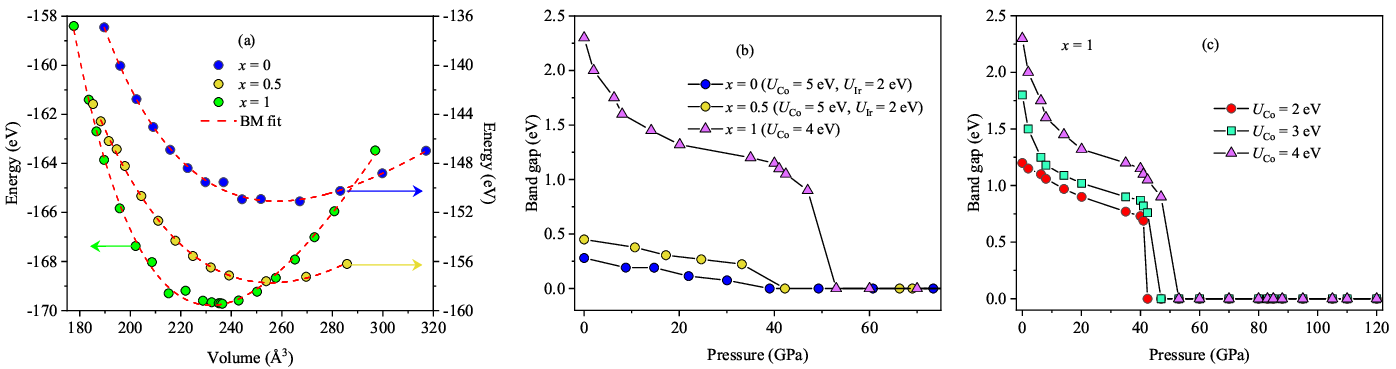}
\caption{(a) Calculated total energy as a function of unit-cell volume for $x$ = 0, 0.5, and 1. The dashed curves represent the corresponding Birch--Murnaghan (BM) equation of state fits used for pressure-dependent band-gap calculations. Variation of the band gap with hydrostatic pressure for (b) $x$ = 0 ($U_\mathrm{Co}$ = 5 eV, $U_\mathrm{Ir}$ = 2 eV), $x$ = 0.5 ($U_\mathrm{Co}$ = 5 eV, $U_\mathrm{Ir}$ = 2 eV), $x$ = 0 ($U_\mathrm{Co}$ = 4 eV), (c) $x$ = 1 at $U_\mathrm{Co}$ = 2~\cite{Nandi2024}, 3, and 4 eV values. It shows a systematic shift of the insulating to metal phase boundaries toward higher values as $U_\mathrm{Co}$ increases.} \label{LCITO_pressure}
\end{figure}

\subsubsection{La$_2$CoIr$_{0.5}$Ti$_{0.5}$O$_6$ ($x$ = 0.5)}

We will now examine the $x$ = 0.5 to understand how partial replacement of Ir by Ti modifies the correlation-dependent structural and electronic behaviour. The $U_{\mathrm{Co}}$ has been varied from 1 to 5 eV in 1 eV steps, with $U_{\mathrm{Ir}}$ fixed at 2 eV. The $x$ = 0.5 system remains metallic at $U_{\mathrm{Co}}$ = 1 eV, whereas a finite band gap of approximately 0.32 eV appears at $U_{\mathrm{Co}}$ = 2 eV. The gap then increases gradually to about 0.45 eV at $U_{\mathrm{Co}}$ = 5 eV, as shown in Fig.~\ref{LCITO}(a), indicating that the insulating state is stabilised as the Co-$3d$ electrons become increasingly localised. The lattice parameters also show noticeable changes between $U_{\mathrm{Co}}$ = 1 and 2 eV, as shown in table~\ref{tab:U_paramters}, followed by only a weak variation at higher $U$ values. This suggests that the initial gap opening might be accompanied by a small structural adjustment, while the subsequent increase in the gap is primarily associated with the electronic correlation. Compared with $x$ = 0, the gap in $x$ = 0.5 opens at a lower value of $U_{\mathrm{Co}}$ and remains larger throughout the investigated range. This suggests that partial replacement of Ir by Ti reduces the contribution of Ir-$5d$ states near the Fermi level and stabilises the insulating state more effectively.

\subsubsection{La$_2$CoTiO$_6$ ($x$ = 1)}

The ground state of $x$ = 1 has been reported~\cite{Nandi2024} as an antiferromagnetic insulator (AFM-I) with a band gap of 1.01 eV within the GGA + $U$ framework at $U_{\mathrm{Co}}$ = 2 eV. In this work, we have extended the study over a wider range of $U_{\mathrm{Co}}$ values from 1 to 5 eV in steps of 1 eV to examine its structural and electronic properties to the correlation strength. Table~\ref{tab:U_paramters} summarises the corresponding lattice parameters and band gap. The lattice parameters remain nearly unchanged with increasing $U_{\mathrm{Co}}$, whereas the band gap increases from 0.53 eV at $U_{\mathrm{Co}}$ = 1 eV to 2.28 eV at $U_{\mathrm{Co}}$ = 5 eV. The increase in band gap is associated with stronger on-site Coulomb interaction and increased localisation of the Co-3$d$ states. Thus, increasing $U_{\mathrm{Co}}$ strengthens the insulating character of $x$ = 1.

\subsection{Application of Hydrostatic Pressure}

\begin{figure}[tbh]
\centering
\includegraphics[width=1.0\linewidth]{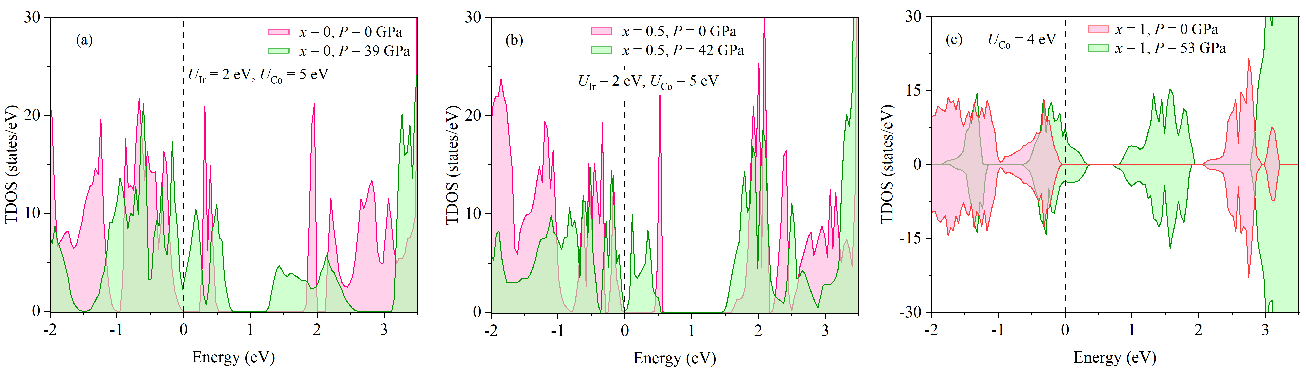}
\caption{(a) Calculated total density of states (TDOS) at ambient pressure and near the critical pressure for the insulator-to-metal transition for (a) $x$ = 0, (b) $x$ = 0.5, and (c) $x$ = 1. For $x$ = 0 and 0.5, the calculations are shown for $U_{\mathrm{Co}}$ = 5 eV and $U_{\mathrm{Ir}}$ = 2 eV. For $x$ = 0, the TDOS is shown at $P$ = 0 and 39 GPa, while for $x$ = 0.5 it is shown at $P$ = 0 and 42 GPa. For $x$ = 1, the TDOS is shown for $U_{\mathrm{Co}}$ = 4 eV at $P$ = 0 and 53 GPa. The dashed vertical line represents the Fermi level ($E_{\mathrm{F}}$ = 0 eV). The appearance of finite states at the Fermi level under pressure indicates the transition from the insulating to the metallic state.} \label{LCITO_dos}
\end{figure}

Hydrostatic pressure provides a useful tuning parameter to modify the electronic state as the calculated band gaps are relatively small and probe the possible insulator to metal transition for $x$ = 0, 0.5, and 1. For this purpose, we calculated the pressure dependence of the band gap for $x$ = 0, 0.5, and 1. For $x$ = 0 and 0.5, the calculations were performed at $U_\mathrm{Co}$ = 5 eV, $U_\mathrm{Ir}$ = 2 eV, and $J_\mathrm{Co}$ = 1.03 eV, $J_\mathrm{Ir}$ = 0.94 eV~\cite{Narayanan2010,Nandi2025}. The energy--volume data were first fitted using the Birch--Murnaghan~\cite{Murnaghan1937,Birch1947} equation of state to describe the compression behaviour of the systems for pressure-dependent calculations, as shown in Fig.~\ref{LCITO_pressure}(a). For $x$ = 0, the band gap decreases progressively with increasing hydrostatic pressure, from about 0.28 eV at ambient pressure to zero near 39 GPa as shown in Fig.~\ref{LCITO_pressure}(b). The TDOS at ambient pressure and 39 GPa, as shown in Fig.~\ref{LCITO_dos}(a), shows the closing of the gap as finite electronic states are present at the Fermi level. For $x$ = 0.5, a similar pressure-driven reduction in the band gap is observed, as presented in Fig.~\ref{LCITO_pressure}(b). The gap decreases from about 0.44 eV at ambient pressure and closes at approximately 42 GPa, indicating a transition from the insulating to the metallic state. The corresponding TDOS at ambient pressure and 42 GPa is shown in Fig.~\ref{LCITO_dos}(b). The finite electronic states present at the Fermi level for 42 GPa reveal the metallic character of the system. For $x$ = 0 and 0.5, the band gap values at ambient pressure are much smaller as compared to $x$ = 1 and changes only slightly at higher $U_\mathrm{Co}$ ($>$ 3 eV) values. Therefore, the critical pressure for gap closing is also expected to change only slightly with further increase in $U_\mathrm{Co}$. The effect of hydrostatic pressure on $x$ = 1 has previously been reported for $U_\mathrm{Co}$ = 2 eV, where an AFM-I to AFM-M transition occurs near 42 GPa~\cite{Nandi2024}. Here, we extend this investigation by applying hydrostatic pressure at $U_\mathrm{Co}$ = 3 and 4 eV to explore how the strength of electron correlation affects the critical pressure for this transition. Figure~\ref{LCITO_pressure}(c) illustrates the variation of band gap for $U_\mathrm{Co}$ = 2, 3, and 4 eV for $x$ = 1. The gap closes at approximately 42, 47, and 53 GPa, respectively, showing that the critical pressure increases with increasing $U_\mathrm{Co}$. The TDOS for $U_\mathrm{Co}$ = 4 eV at ambient pressure and 53 GPa is shown in Fig.~\ref{LCITO_dos}(c). The closing of the band gap under pressure reflects an increase in bandwidth and orbital overlap, which reduces the correlation-induced gap.

\section{Summary}

For the La$_2$CoIr$_{1-x}$Ti$_x$O$_6$ series, we have systematically investigated the structural and electronic properties for $x$ = 0, 0.5, and 1 using first-principles electronic structure calculations. The effect of electron correlation was examined by varying $U_{\mathrm{Co}}$, while the role of the Ir site was further studied for $x$ = 0 by changing $U_{\mathrm{Ir}}$ at fixed $U_{\mathrm{Co}}$. The results show that increasing $U$ generally enhances the insulating character, while the lattice parameters change only slightly. For $x$ = 0, the band gap is found to be more sensitive to $U_{\mathrm{Ir}}$ than to $U_{\mathrm{Co}}$. Hydrostatic pressure was then applied to all three compositions, showing a gradual reduction of the band gap and pressure-induced insulator-to-metal transitions. Overall, this study highlights that across the La$_2$CoIr$_{1-x}$Ti$_x$O$_6$ series, increasing $U$ generally strengthens the insulating state by increasing the band gap, whereas hydrostatic pressure drives the opposite trend by reducing the gap and eventually inducing a metallic state. This contrasting response shows that the electronic state is controlled by the competition between electron correlation and bandwidth broadening due to hydrostatic pressure. The results also indicate that the critical pressure for the insulator-to-metal transition depends on the strength of electron correlation, as large $U$ values require higher pressure values to drive the system into a metallic state.

\section*{Acknowledgments}

RSM acknowledges the financial support from SERB for his Early Career Research Award with file number ECR/2018/000999/PMS. The authors would like to acknowledge IIT Tirupati for providing the high-performance computing (HPC) cluster for DFT calculations.


\begin{thebibliography}{9}

\bibitem{Vasala2015} S. Vasala and M. Karppinen, Prog. Solid State Chem. \textbf{43}, 1 (2015).

\bibitem{Cavichini2018} A. S. Cavichini, M. T. Orlando, J. B. Depianti, J. L. Passamai, Jr., F. Damay, F. Porcher, and E. Granado, Phys. Rev. B \textbf{97}, 054431 (2018).

\bibitem{Yasir2025} Md. Y. Arafat, S. Akter, Md. F. Shanto, M. J. Hosen, and  M. D. I. Bhuyan, Phys. Chem. Chem. Phys. \textbf{27}, 4652 (2025).

\bibitem{Nandi2024} S. Nandi, S. Pradhan, A. K. Nandy, and R. S. Manna, Phys. Rev. B \textbf{110}, 085101 (2024).

\bibitem{Kresse1996} G. Kresse and J. Furthm\"{u}ller, Phys. Rev. B \textbf{54}, 11169 (1996).

\bibitem{Blochl1994} P. E. Bl\"{o}chl, Phys. Rev. B \textbf{50}, 17953 (1994).

\bibitem{Kresse1999} G. Kresse and D. Joubert, Phys. Rev. B \textbf{59}, 1758 (1999).

\bibitem{PBE1996} J. P. Perdew, K. Burke, and M. Ernzerhof, Phys. Rev. Lett. \textbf{77}, 3865 (1996).

\bibitem{Nandi2025} S. Nandi, V. Yadav, S. Devi, C. S. Yadav, B. Das, S. Datta, K. Dolui, and R. S. Manna, Phys. Rev. B \textbf{112}, 155144 (2025).

\bibitem{Mei2009} Z.-G. Mei, S. Shang, Y. Wang, and Z.-K. Liu, Phys. Rev. B \textbf{79}, 134102 (2009).

\bibitem{Narayanan2010} N. Narayanan, D. Mikhailova, A. Senyshyn, D. M. Trots, R. Laskowski, P. Blaha, K. Schwarz, H. Fuess, and H. Ehrenberg, Phys. Rev. B \textbf{82}, 024403 (2010).

\bibitem{Murnaghan1937} F. D. Murnaghan, Am. J. Math. \textbf{59}, 235 (1937).

\bibitem{Birch1947} F. Birch,  Phys. Rev. \textbf{71}, 809 (1947).

%\bibitem{Rodriguez2002} E. Rodr\'{i}guez, M. L. L\'{o}pez, J. Campo, M. L. Veiga, and C. Pico, J. Mater. Chem. \textbf{12}, 2798 (2002).

%\bibitem{Dudarev1998} S. L. Dudarev, G. A. Botton, S. Y. Savrasov, C. J. Humphreys, and A. P. Sutton, Phys. Rev. B \textbf{57}, 1505 (1998).

%\bibitem{Sirothiya2019} D. M. Sirothiya, N. Solanki, N. Kaurav, and A. Mishra, AIP Conf. Proc. \textbf{2100}, 020139 (2019).

\end{thebibliography}
\end{document}